# Spin valve effect in junctions with a single ferromagnet

Fengrui Yao[1,2*], Volodymyr Multian[1,2,3], Kenji Watanabe[4], Takashi Taniguchi[5], Ignacio Gutiérrez-Lezama[1,2], and Alberto F. Morpurgo[1,2*]

[1]*Department of Quantum Matter Physics, University of Geneva, 24 Quai Ernest Ansermet, CH-1211 Geneva, Switzerland*

[2]*Group of Applied Physics, University of Geneva, 24 Quai Ernest Ansermet, CH-1211 Geneva, Switzerland*

[3]*Advanced Materials Nonlinear Optical Diagnostics lab, Institute of Physics, NAS of Ukraine, 46 Nauky pr., 03028 Kyiv, Ukraine*

[4]*Research Center for Electronic and Optical Materials, National Institute for Materials Science, 1-1 Namiki, Tsukuba, 305-0044, Japan*

[5]*Research Center for Materials Nanoarchitectonics, National Institute for Materials Science, 1-1 Namiki, Tsukuba, 305-0044, Japan*

*Correspondence: fengrui.yao@unige.ch; alberto.morpurgo@unige.ch

**Abstract**

Spin valves are essential components in spintronic memory devices, whose conductance is modulated by controlling spin-polarized electron tunneling through the alignment of the magnetization in ferromagnetic elements. Whereas conventional spin valves unavoidably require at least two ferromagnetic elements, here we demonstrate a van der Waals spin valve based on a tunnel junction that comprises only one such ferromagnetic layer. Our devices combine a $Fe_3GeTe_2$ electrode acting as spin injector together with a paramagnetic tunnel barrier, formed by a $CrBr_3$ multilayer operated above its Curie temperature. We show that these devices exhibit a conductance modulation with values comparable to that of conventional spin valves. A quantitative analysis of the magnetoconductance that accounts for the field-induced magnetization of $CrBr_3$, and that includes the effect of exchange interaction, confirms that the spin valve effect originates from the paramagnetic response of the barrier, in the absence of spontaneous magnetization in $CrBr_3$.

**Key word:** Spin valve effect; Tunnelling junction; Magnetic properties; Brillouin function

**Introduction**

Spin valves – the building blocks of magnetoresistive random-access memories – are devices whose conductance depends on the relative alignment of the magnetization in two ferromagnetic elements[1-6]. The most common form of spin-valve devices consists of two ferromagnetic metals separated by a non-magnetic insulating barrier[7-17] (Fig. 1a). The imbalance in the density of states of the majority and minority spins in the ferromagnetic electrodes causes the tunnelling conductance to be larger or smaller, depending on whether when the magnetization of the two electrodes are parallel or antiparallel. A less common form of spin-valve structure combines a single ferromagnetic electrode with an insulating ferromagnetic barrier acting as spin filter[18-25] (a second non-magnetic electrode is also present; see Fig. 1b). In simple terms, the device conductance is determined by the height of the tunnel barrier, which in a magnetic insulator depends on spin, so that – when the magnetization of the ferromagnetic electrode and of the barrier are parallel – majority spin electrons are transmitted with higher probability, and the conductance is larger than for antiparallel magnetizations. In principle, a third type of spin valve could be realized, based on a tunnel barrier with two ferromagnetic insulators (and non-magnetic metallic electrodes), whose relative magnetization orientations determines the probability of electron tunnelling. Devices comprising A-type antiferromagnetic insulating bilayers can be viewed as a practical realization of this third type of device structure[26-30].

What is common to all devices just described is that two distinct ferromagnetic elements are required to observe a spin-valve effect. Such a requirements seems trivially obvious, because in a spin valve it is the relative orientation of the magnetizations in the two ferromagnetic elements that determines the device resistance. Devices with a single ferromagnetic element can exhibit finite magnetoconductance due to the coupling of an applied magnetic field to the magnetization, due –for instance– to anisotropic magnetoresistance effects[31, 32] (i.e., the dependence of the transport properties of a ferromagnet on the orientation of the magnetization relative to the crystalline axis of a material). To this day, however, no device with a single ferromagnetic element has been reported to exhibit magnetoconductance originating from modulation of the transport properties due to a spin-valve effect.

Despite this seemingly obvious knowledge, here we show experimentally that a sizable spin-valve magnetoconductance can be observed in devices containing only one individual ferromagnetic element, namely a tunnel barrier with one ferromagnetic metallic contact. The tunnel barrier that we consider is formed by a paramagnetic insulator with an extremely large magnetic susceptibility. Because of the very large susceptibility, the magnetic field applied to switch the magnetization of the ferromagnetic electrode induces a finite magnetization in the barrier. It is this field-induced magnetization that – in the absence of a spontaneous magnetization– is responsible for the appearance of a spin-valve effect. Contrary to the case of the conventional devices described above, which commonly exhibit two switching fields, associated to the flipping of the magnetization in the two ferromagnetic elements, the spin-valve effect that we observe increases gradually in magnitude upon increasing the applied field from $H = 0$ T, and exhibits only one switching field (corresponding to the coercive field of the metallic ferromagnet employed as electrode.

Our strategy to demonstrate the concept of a spin-valve device with a single ferromagnetic contact consists in selecting a ferromagnetic insulator for the barrier, and operate the device above its Curie temperature $T_c$, i.e., in a temperature range where the barrier material is paramagnetic. Not only does this strategy makes it straightforward to achieve experimentally a very large magnetic susceptibility $\chi$ –since as $T_c$ is approached from the paramagnetic state $\chi$ tends to diverge[33]– but also it allows testing the response of the spin-valve upon varying $\chi$, which can be done simply by varying the temperature. For these reasons, we realize spin-valve devices that employ $Fe_3GeTe_2$ (FGT) as metallic ferromagnetic electrode and $CrBr_3$ as barrier material. Such a choice is ideal for a few reasons. First, the critical temperature of FGT[13, 34-37] (200 K in our crystals, see Supporting Information Fig.S1 ) is much larger than that of $CrBr_3$[25, 38-41] (an easy axis ferromagnet with $T_c \sim$ 30-32 K), which allows measuring the devices in a broad temperature range above $T_c$ of $CrBr_3$, without modifying substantially the properties of FGT. Additionally, the relatively large coercive field of FGT (approximately 400 mT in our devices, much larger than that of $CrBr_3$ in the ferromagnetic state) enables a very large spin-polarization to be induced in the paramagnetic state of $CrBr_3$ as $T$ approaches $T_c$, such that the magnitude of the spin-valve signal in the paramagnetic state is comparable to that in the ferromagnetic state. This facilitates enormously the quantitative analysis of our data.

Devices are fabricated by exfoliating FGT and $CrBr_3$ bulk crystals in a glove-box and by using established techniques to pick layers and transfer them to form the desired structure[42], as shown in the schematics in the top inset of Fig. 1c (the hBN encapsulating layers needed to prevent degradation in air are not shown; see Supporting Information Method section and Fig.S2 for details of the device fabrication). The FGT layers are typically 5 to 10 nm thick, while the thickness of $CrBr_3$ ranges typically between 3.5 and 8.5 nm. We have realized and investigated four different devices, all exhibiting virtually identical behavior, and analyzed quantitatively in details two of them. The data shown in the main text and Supporting Information Fig.S3 and Fig.S4 have all been measured on one of these devices (with $CrBr_3$ thickness of 3.5 nm); data from the second device (with $CrBr_3$ thickness of 8.5 nm) analyzed in detail are shown in Supporting Information Fig.S5 and Fig.S6.

Fig. 1c shows the *I-V* characteristics of one of our junctions –indicative of Fowler-Nordheim tunnelling[43, 44] (see bottom inset). To ensure the proper operation of our devices, we start by characterizing their magnetoconductance at low temperature ($T = 2$ K), when both the FGT electrode and the CrBr$_3$ barrier are ferromagnetic, so that a conventional spin-valve effect is expected. Fig. 2a shows the device magnetoconductance $\delta G(H, T) = (G(H,T) - G_0(T))/G_0(T)$ ($G(H,T)$ is the conductance measured at temperature $T$ and magnetic field $H$ and $G_0(T) = G(H = 0,T)$) measured at $T = 2$ K, upon sweeping the magnetic field up ($\delta G_\uparrow$; blue trace) or down ($\delta G_\downarrow$; red trace). Upon injecting electrons from the FGT contact, a hysteretic magnetoconductance is observed as expected for a spin valve, superimposed on a background independent of the injecting electrode (i.e., the background magnetoconductance is the same when injecting from the graphene electrode). The background magnetoconductance ($\delta G_{bg} = ((\delta G_\uparrow + \delta G_\downarrow) + (|\delta G_\uparrow - \delta G_\downarrow|))/2$) originates from moiré magnetism in the tunnel barrier, and it is observed in CrBr$_3$ barriers with FGT contacts (i.e., no background is observed in CrBr$_3$ barriers with graphene contacts). As we reported very recently[41], that is because the different thermal expansion coefficients of FGT and CrBr$_3$ cause differential strain inside the CrBr$_3$ barrier (i.e., a small difference in lattice constant between individual CrBr$_3$ layers) which results in the appearance of a moiré pattern (in the absence of any twisting) with accompanying moiré magnetism. Details on moiré magnetism can be found in Ref. 41, as the present manuscript is devoted exclusively to the behavior of the spin-valve magnetoconductance, with a particular focus on its evolution for $T > T_c$ (which we did not discuss in Ref. 41).

Fig. 2b shows the spin-valve contribution to the magnetoconductance measured upon sweeping the applied magnetic field up (top panel) or down (bottom panel). the magnetoconductance switches as the applied magnetic field passes through zero, and switches back once the field is swept past the coercive field of FGT at approximately 400 mT. As the coercive field of CrBr$_3$ is much smaller than that of FGT, its precise determination requires the measurement of "minority" hysteresis loops, in which the sweeping direction of the applied magnetic field is reversed before exceeding the coercive field of FGT. The corresponding data are shown in Fig. 2c, from which we estimate the coercive field of CrBr$_3$ to be approximately 40 mT at low temperature. Similar minority loop measurements performed at different temperatures allow a first determination of the $T_c$ of CrBr$_3$, simply by identifying the temperature at which hysteresis disappears (due to the vanishing coercive field of CrBr$_3$). The Curie temperature of CrBr$_3$ determined in this way is found to be between 30 and 32 K, these values are in close agreement with the ones determined precisely by collapsing the magnetoconductance curves in the paramagnetic phase of CrBr$_3$, by plotting them as a function of $\frac{H}{T-T_c}$, which near $T_c$ is proportional to the magnetization[25] (see Supporting Information Fig.S3). specially, we find that $T_c = 31$ K for the device discussed in the main text, and $T_c = 32$ K for the device whose data are shown in Supporting Information, in close agreement with literature values[25]. In summary, when CrBr$_3$ is in its ferromagnetic state the device response conforms to expectations: it exhibits the expected hysteretic magnetoconductance (larger than 15%), with two sharp switches at the coercive field of CrBr$_3$ and FGT, with the hysteresis that disappears when the temperature is increased past the Curie temperature of CrBr$_3$, i.e., when CrBr$_3$ becomes paramagnetic.

Even if the magnetoconductance hysteresis in the minority loops disappears as $T$ is increased past $T_c$ and the CrBr$_3$ coercive field vanishes, the magnetoconductance measured over a magnetic field interval larger than the FGT coercive field continues to exhibit hysteresis (see color plot in Fig. 3a and Supporting Information Fig.S4). This observation comes as a surprise, because It indicates that a sizable spin-valve effect persists when CrBr$_3$ is paramagnetic. To appreciate the difference in the behavior of the spin-valve magnetoconductance with CrBr$_3$ either ferromagnetic or paramagnetic, we plot in Fig. 3b and 3c cuts of the color plot of Fig. 3a for $T$ respectively lower or higher than the CrBr$_3$ critical temperature. It is apparent that the nature of the transition from negative to positive $(\delta G_\downarrow - \delta G_\uparrow)$ is different depending on whether $T$ is below or above $T_c$. If $T < T_c$, the transition is abrupt, as expected from the switch of the CrBr$_3$ magnetization; if $T > T_c$, the transition is gradual. This difference is illustrated by the color plot of $\frac{d(\delta G_\downarrow - \delta G_\uparrow)}{dH}$ (Fig. 3d), which highlights the abrupt change in the sharpness of the transition $(\delta G_\uparrow - \delta G_\downarrow)$ from negative to positive as $T$ crosses $T_c$ (looking at the $T$ dependence of $\frac{d(\delta G_\downarrow - \delta G_\uparrow)}{dH}$ provides one more estimate of $T_c$ in CrBr$_3$).

The different evolution of $(\delta G_\downarrow - \delta G_\uparrow)$ with applied magnetic field discussed here provides an indication of the origin of the spin valve effect when CrBr$_3$ is paramagnetic. When $T > T_c$, the magnetization $M$ of CrBr$_3$ vanishes at zero applied magnetic field $\mu_0 H = 0$ T. However, because of the very large magnetic susceptibility, the application of even a small magnetic field creates a non-negligible magnetization parallel to the applied field (Fig.4a). It is this field-induced magnetization that generates a spin valve effect in the absence of a spontaneous magnetization, and that accounts for the magnetoconductance observed for $T > T_c$. The field-induced magnetization also leads to a rather large hysteresis, because the field needed to switch the magnetization in the FGT electrode is sizable. Therefore, as the magnetic field is swept past the coercive field of FGT, the direction of the magnetization in the FGT electrode switches and becomes parallel to the applied field, generating an abrupt change in the junction magnetoconductance.

The scenario just proposed –namely that the observed magnetoconductance is due to the spin valve effect associated to the field-induced magnetization in the paramagnetic state of CrBr$_3$– can be validated quantitatively by estimating at the mean field level how the magnetization of CrBr$_3$ evolves with applied magnetic field. For a paramagnet formed by independent spins with $S = 3/2$ ($S = 3/2$ corresponds to the case of Cr atoms), the magnetization is given by[45]

$$M = M_{\text{Sat}} B_S \left(g\mu_B S \frac{H}{k_B T}\right), \tag{1}$$

where $M_{\text{Sat}} = n\, g\mu_B S$, with $n$ density of spins in the material, $g$ the gyromagnetic ratio, $\mu_B$ the Bohr magneton, and $B_S(x) = \frac{2S+1}{2S}\coth\left(\frac{2S+1}{2S}x\right) - \frac{1}{2S}\coth\left(\frac{1}{2S}x\right)$ is the Brillouin function (here $x = g\mu_B S \frac{H}{k_B T}$, is the ratio of the Zeeman energy of the magnetic moment in the external field to the thermal energy $k_B T$). In the presence of exchange interaction, the expression for the magnetization at mean field level is modified into[45]:

$$M = M_{\text{Sat}} B_S \left(g\mu_B S \frac{H_{\text{eff}}}{k_B T}\right), \tag{2}$$

with $H_{\text{eff}} = H + \lambda M$, and $\lambda$ a constant that quantifies the strength of the exchange interaction. For $T > T_c$, $M = \chi H$ in the linear regime, and to a very good approximation we have:

$$M = M_{\text{Sat}} B_S \left(g\mu_B S \frac{H(1+\lambda\chi)}{k_B T}\right). \tag{3}$$

To employ these formulas in the analysis of the experimental data it suffices to linearize Eq. (3) for $T$ near $T_c$, and consider that $\chi \propto \frac{1}{T-T_C} \gg 1$. Because the spin-valve magnetoconductance ($\delta G_\uparrow - \delta G_\downarrow$) is proportional to the magnetization $M$ [25], it follows that for $T$ near $T_c$ the slope $\left[\frac{d(\delta G_\uparrow - \delta G_\downarrow)}{dH}\right]_{H=0} \propto \frac{1}{T-T_C}$ (top panel of Fig. 4c). It also follows that the full magnetoconductance curve is well approximated by the Brillouin function, exhibiting a non-linearity that becomes increasingly more pronounced as $T$ approaches $T_c$ (because the term $\lambda\chi$ diverges proportionally to $\frac{1}{T-T_C}$, since $\lambda$ is constant).

A direct comparison with measurements shows that these conclusions are indeed in excellent agreement with experimental data. The top panel of Fig. 4c shows that the slope $\left[\frac{d(\delta G_\uparrow - \delta G_\downarrow)}{dH}\right]_{H=0}$ is proportional to $\frac{1}{T-T_C}$ throughout the investigated range (i.e., for $T$ up to 50 K). Even nearly 20 K above the Curie temperature –deep in the paramagnetic phase of $CrBr_3$– the magnetoconductance due to the spin-valve effect of the field-induced magnetization is easily detectable. The non-linearity of the Brillouin function can only be detected when $T$ is sufficiently close to $T_c$ (for $T$ smaller than 38 K), because at larger temperature the spin-valve magnetoconductance is virtually linear on the scale of the coercive field of FGT. Within this smaller temperature range, we fit the functional dependence of the magnetoconductance to that of the magnetization given in Eq. (3), using $\lambda\chi$ as fitting parameter, and plot the results for the best value obtained for the fitting parameter (bottom panel of Fig. 4c). Albeit limited, the temperature range is still sufficient to exhibit the expected proportionality of $\lambda\chi$ to $\frac{1}{T-T_C}$:

indeed, since $\lambda$ is constant the temperature dependence is determined exclusively by $\chi$. The red dashed curves in the different panels of Fig. 4b illustrate the quality of the fits, and confirm that throughout the experimentally accessible temperature and magnetic field range, the hysteretic contribution to the magnetoconductance ($\delta G_\downarrow - \delta G_\uparrow$) scales proportionally to the calculated field-induced magnetization (for the same analysis for the device with 8.5 nm $CrBr_3$ barrier are shown in Supporting Information Fig.S5 and Fig.S6).

Besides showing that a fairly complete understanding of magnetotransport can be obtained from basic concepts, without the need of additional assumptions, establishing that the temperature and magnetic field dependence of spin valve magnetoconductance conforms quantitatively to the behavior expected for a paramagnet is important for one more reason. It excludes the possibility that the spin valve effect originates from an interfacial layer of $CrBr_3$ that remains ferromagnetic at temperatures higher than the Curie temperature of the $CrBr_3$ multilayer, because of the proximity with a stronger ferromagnet such as FGT (whose Curie temperature is 200 K). From our measurements and their quantitative analysis, we can therefore conclude that a finite –and relatively large– spin valve magnetoconductance can be observed also in devices with a single ferromagnetic element.

The observation of spin valve magnetoconductance in a structure formed by a single ferromagnet is unexpected, since so far spin valve effects have been documented only in devices with two distinct ferromagnetic elements. Our analysis explains the origin of the phenomenon in terms of field-induced magnetization, since in devices consisting of a ferromagnetic electrode and a paramagnetic tunnel barrier, the spontaneous magnetization in the barrier vanishes. However, owing to the magnetic susceptibility of the paramagnetic state, when an external magnetic field is applied a finite magnetization is nevertheless present. This is all what is needed to generate a spin valve magnetoconductance. The concept is more general that the specific realization demonstrated in this work. Indeed, we can easily envision a spin valve with a ferromagnetic and a paramagnetic metal separated by a non-magnetic tunnel barrier, generating a spin-valve effect similarly to what we have shown here. as the paramagnetic metal is magnetized by the applied magnetic field, spin valve magnetoconductance can result from the relative orientation of the spontaneous magnetization in the ferromagnetic electrode and the field-induced magnetization in the paramagnetic one. Conceptually, choosing as paramagnet a ferromagnet operated at temperatures $T > T_c$ is not necessary. Nevertheless, such a choice clearly helps to achieve large effects –when measurements are done for $T$ near $T_c$– owing to the divergence of the magnetic susceptibility. In that case, selecting a material with $T_c$ much lower than the actual ferromagnetic electrode and with a much smaller coercive field, which is what we have done here, provides ideal conditions to observe and analyze the effect in a broad range of temperature and applied magnetic fields.

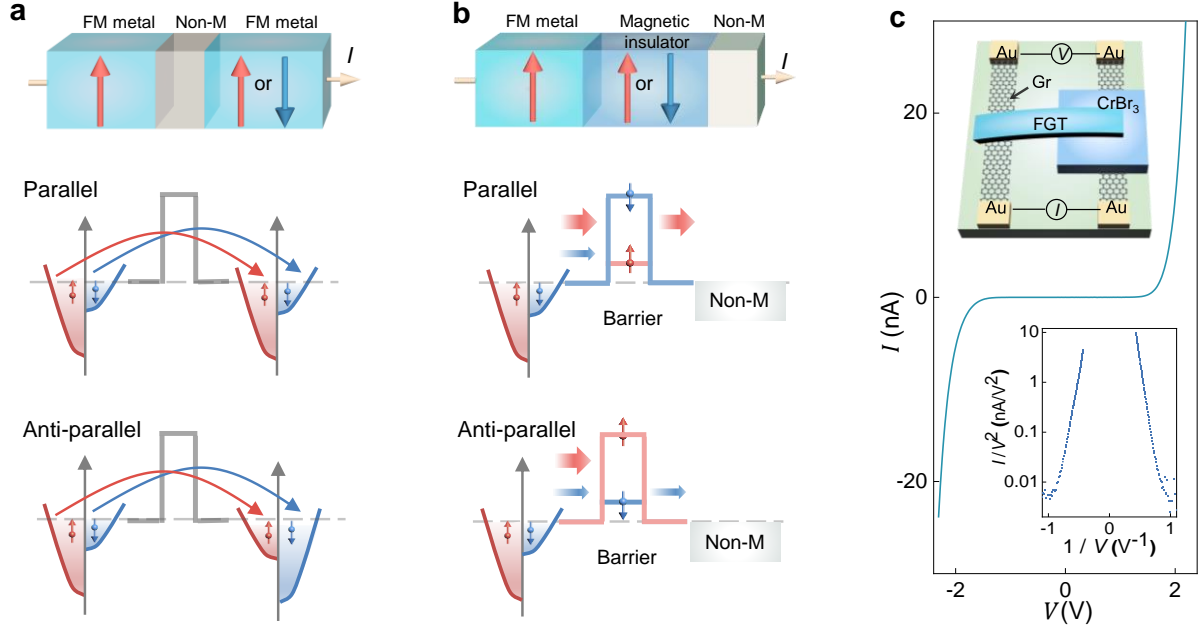

**Figure 1: Conventional spin-valve architectures. a,** Conventional spin valve based on ferromagnetic electrodes separated by a non-magnetic insulator (top panel). The tunnelling conductance is determined by the imbalance in the density of states of majority and minority carriers in the ferromagnets. When the applied magnetic field aligns the magnetization in the ferromagnetic electrodes the tunnelling conductance is large (parallel configuration; middle panel), whereas the anti-parallel configuration (bottom panel) results in low conductance. **b,** Spin valve based on a ferromagnetic insulator placed in between a ferromagnetic and a non-magnetic electrode (top panel). The magnetic insulator acts as a spin filter, with the height of the barrier for the injected majority spins determining the tunnelling conductance. When the applied magnetic field causes an anti-parallel alignment of the magnetizations, the tunnel barrier for majority spins is high (middle panel) and the tunnelling conductance is low; parallel magnetizations (bottom panel) result in a low barrier height and a high tunnelling conductance. **c,** Current-voltage (*I-V*) characteristics of our $Fe_3GeTe_2$(FGT)/$CrBr_3$/graphene (Gr) tunnel junction ($T = 2$ K; see top inset for device schematics). The bottom inset shows that log ($I/V^2$) is proportional to $1/V$ over several decades, indicating that tunnelling across $CrBr_3$ occurs in the Fowler-Nordheim regime.

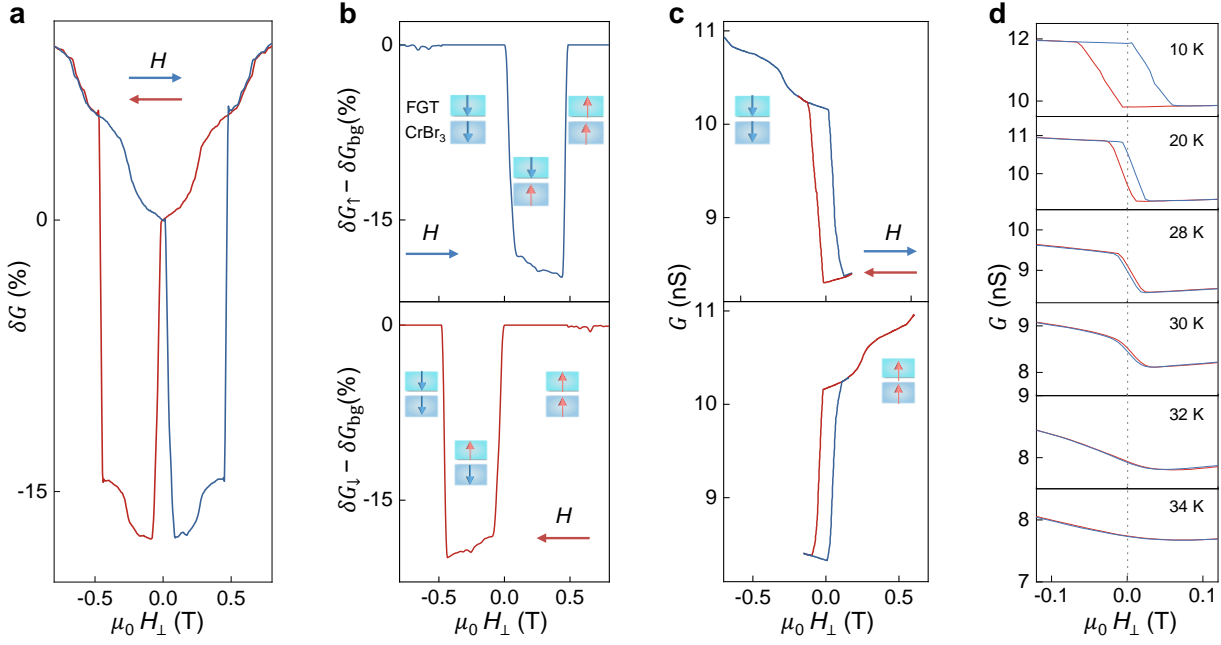

**Figure 2: Low-temperature spin-valve effect in FGT/CrBr₃/Gr. a,** When FGT and CrBr$_3$ are both in the ferromagnetic state the tunneling magnetoconductance $\delta G(H, 2K)$, measured with applied bias $V = 1.2V$, exhibits the hysteresis characteristic of spin valves, superimposed on a background ($\delta G_{bg} = ((\delta G_\uparrow + \delta G_\downarrow) + (|\delta G_\uparrow - \delta G_\downarrow|))/2$) originating from differential strain in CrBr$_3$[41]. **b,** Background-subtracted magnetoconductance measured upon sweeping up ($\delta G_\uparrow - \delta G_{bg}$; top panel) or down ($\delta G_\uparrow - \delta G_{bg}$; bottom panel) the applied magnetic field. The switching fields separating the low and high conductance correspond well to the coercive fields of CrBr$_3$ (40 mT) and FGT (400 mT). **c,** Minority magnetoconductance loops enable the precise determination of the coercive field of CrBr$_3$. Top (bottom) panel: after applying a field of -0.6T (0.6T), for which the magnetizations of FGT and CrBr$_3$ both point downward (upwards), the field is swept past zero and the sweep direction is reversed at 150 mT (-150 mT; before reaching the coercive field of FGT). **d,** Temperature-evolution of the minority hysteresis loops. The disappearance of magnetic hysteresis in the minority loops between 30 K and 32 K gives a first estimate of the Curie temperature of CrBr$_3$. In all panels the magnetic field has been applied perpendicularly to the layers ($H_\perp$), and the blue (red) trace correspond to sweeping up (down) the magnetic field (as indicated by the arrow).

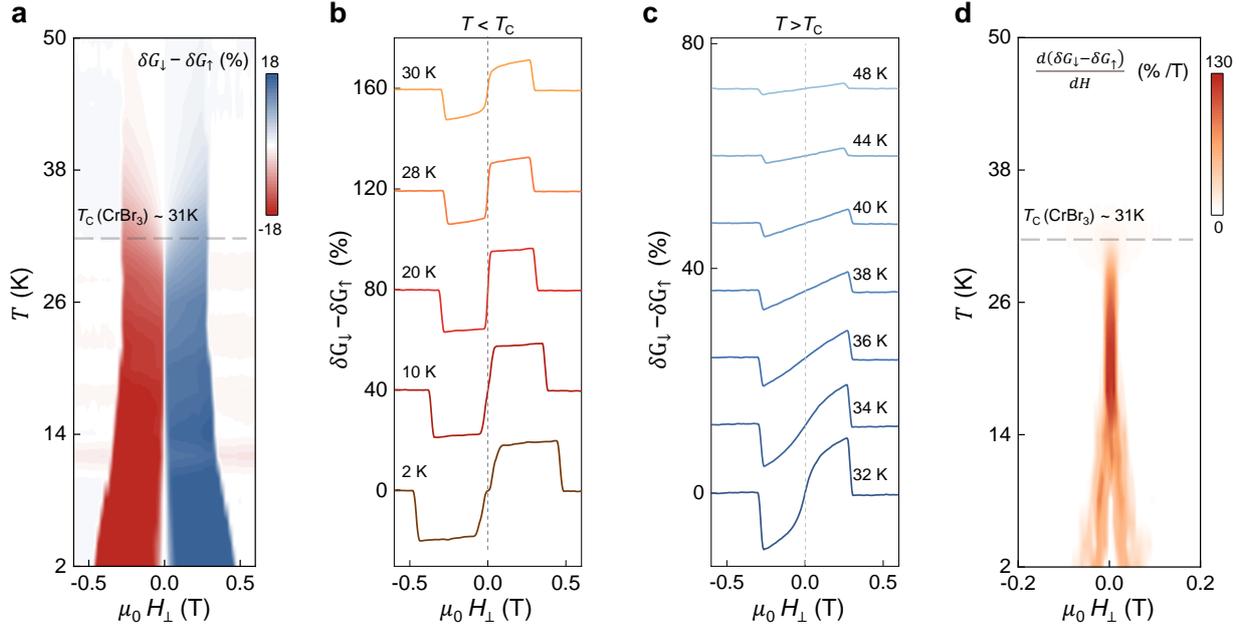

**Figure 3: Persistence of spin valve effect above the Curie temperature of CrBr₃.** The hysteresis due to the spin valve effect manifests itself in a finite difference between the magnetoconductance measured upon sweeping the magnetic field up or down, $\delta G_\downarrow - \delta G_\uparrow$. **a,** The color map of $\delta G_\downarrow - \delta G_\uparrow$ shows unambiguously a spin valve effect persisting at temperatures well above $T_c$ of CrBr₃ (~31K). **b-c,** The dependence of $\delta G_\downarrow - \delta G_\uparrow$ on magnetic field ($H_\perp$, applied perpendicularly to the layers) is different for $T < T_c$ (**b**) and $T > T_c$ (**c**). Below $T_c$, a sharp transition is observed near zero field (between ±40 mT at low $T$), due to the finite coercive field of CrBr₃. Above $T_c$, $\delta G_\downarrow - \delta G_\uparrow$ exhibits a continuous increase on a larger field scale (~ ± 250 mT). The difference is apparent in the color map of the slope $d(\delta G_\downarrow - \delta G_\uparrow)/dH_\perp$ plotted as a function of magnetic field and temperature (panel **d**), which illustrates clearly how the magnitude of the slope changes abruptly when $T$ is increased above $T_c$ ~31K. The change is caused by the transition of CrBr₃ from the ferromagnetic to the paramagnetic state, since for $T > T_c$ (i.e., in the paramagnetic state) the coercive field of CrBr₃ vanishes, no spontaneous magnetization persists, and the magnetization in CrBr₃ is exclusively field induced.

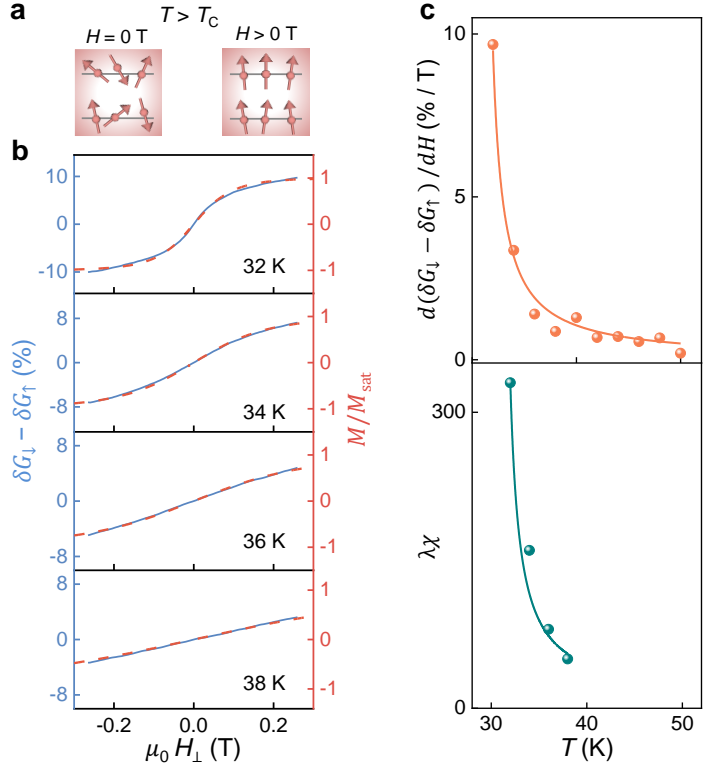

**Figure 4: Origin of the spin valve effect above the Curie temperature of CrBr$_3$. a,** Schematic illustration of the field-induced magnetization $M$ of CrBr$_3$ above its Curie temperature. Above $T_c$ and in the absence of an external magnetic field, CrBr$_3$ exhibits zero net magnetization, but due to the large susceptibility of CrBr$_3$, the application of even a small magnetic field results in a finite magnetization parallel to the applied field. It is this field-induced magnetization that is responsible for the observation of the spin valve effect above the $T_c$ of CrBr$_3$. **b,** Comparison between the spin-valve magnetoconductance $\delta G_\downarrow - \delta G_\uparrow$ (blue line) and the magnetization $M/M_{sat}$ (red dashed line; $M_{sat}$ is the saturation magnetization) calculated using the Brillouin function as described in the main text (at each temperature, the quantity $\lambda\chi$ is used as fitting parameter). The good agreement demonstrates that for $T > T_c$ the hysteretic contribution to the magnetoconductance is proportional to the field-induced magnetization of CrBr$_3$. **c.** Both the zero field slope of the spin-valve magnetoconductance ($[\frac{d(\delta G_\uparrow - \delta G_\downarrow)}{dH}]_{H=0}$, top panel) and the fitting parameter $\lambda\chi$ (bottom panel) extracted from fit of the spin-valve magnetoconductance shown in panel **b**, scale proportionally to $1/(T-T_c)$ (continuous orange and cyan lines in the top and bottom panels are fits to the data points), as expected from Weiss law.

**Supporting Information**

Methods; Anomalous Hall effect of $Fe_3GeTe_2$ thin layers; Optical micrographs of devices; Temperature-dependent spin-valve magnetoconductance in additional devices

**Acknowledgments**

The authors gratefully acknowledge Alexandre Ferreira for technical support and useful discussions with Zhe Wang. A. F. M. gratefully acknowledges the Swiss National Science Foundation (Division II, project #200021_219424) and the EU Graphene Flagship project for support. K.W. and T.T. acknowledge support from the JSPS KAKENHI (Grant Numbers 21H05233 and 23H02052), the CREST (JPMJCR24A5) and World Premier International Research Center Initiative (WPI), MEXT, Japan.

**TOC graphic**

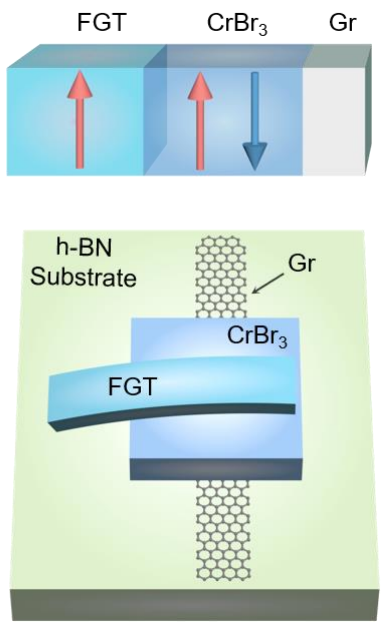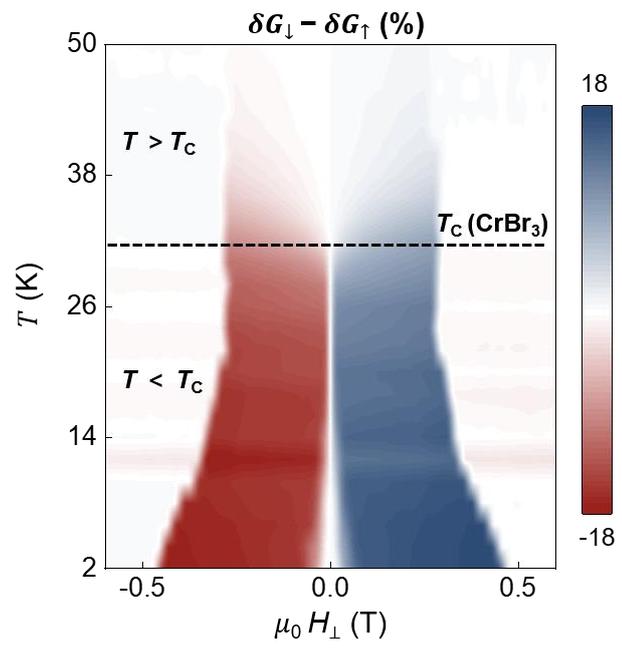